\documentclass[11pt]{article}
\pdfoutput=1
\usepackage{setspace}

\usepackage[letterpaper,margin=1in]{geometry}
\usepackage[backref,colorlinks,citecolor=blue,bookmarks=true]{hyperref}

\usepackage{amsmath,amsfonts,amssymb,bbm} 
\usepackage[numbers,sort&compress]{natbib} 

\usepackage{forest}
\usepackage{standalone}

\usepackage[numbers]{natbib}

\newcount\Comments  
\definecolor{thegreen}{rgb}{0,0.6,0.1}
\definecolor{thered}{rgb}{0.8,0.2,0.1}
\definecolor{teal}{rgb}{0.1,0.6,0.6}
\newcommand{\kibitz}[2]{\ifnum\Comments=1\textcolor{#1}{#2}\fi}

\newcommand{\figreduc}{\vspace{-0mm}}
\newcommand{\efg}{EFG}

\usepackage{paralist}
\usepackage{mathtools}
\usepackage{wrapfig}
\usepackage{amsmath}
\usepackage{amssymb}
\usepackage{amsthm}
\usepackage{hyperref}
\usepackage{url}

\theoremstyle{plain}
\newtheorem{theorem}{\protect\theoremname}
\theoremstyle{definition}
\newtheorem{definition}{\protect\definitionname}
\theoremstyle{plain}

\theoremstyle{plain}
\newtheorem{proposition}{\protect\propositionname}
\providecommand{\definitionname}{Definition}
\providecommand{\lemmaname}{Lemma}
\providecommand{\propositionname}{Proposition}
\providecommand{\theoremname}{Theorem}

\newcommand{\tuple}[1]{\ensuremath{\left \langle #1 \right \rangle }}
\newcommand{\child}[2]{\ensuremath{t^{#1}_{#2}}}
\newcommand{\gameclass}{CRSWF}
\newcommand{\gameclassfull}{chance-relaxed skew well-formed}
\newcommand{\Gameclassfull}{Chance-relaxed skew well-formed}

\begin{document}

\markboth{Anonymous}{Imperfect-Recall Abstractions With Bounds}

\title{Imperfect-Recall Abstractions with Bounds in Games}

\author{
Christian Kroer and Tuomas Sandhom\\
Computer Science Department\\
Carnegie Mellon University
}

\maketitle

%



\begin{abstract}
  Imperfect-recall abstraction has emerged as the leading paradigm for practical large-scale equilibrium computation in imperfect-information games. However, imperfect-recall abstractions are poorly understood, and only weak algorithm-specific guarantees on solution quality are known. We develop the first general, algorithm-agnostic, solution quality guarantees for Nash equilibria and approximate self-trembling equilibria computed in imperfect-recall abstractions, when implemented in the original (perfect-recall) game. Our results are for a class of games that generalizes the only previously known class of imperfect-recall abstractions for which any such results have been obtained. Further, our analysis is tighter in two ways, each of which can lead to an exponential reduction in the solution quality error bound.

  We then show that for extensive-form games that satisfy certain properties, the problem of computing a bound-minimizing abstraction for a single level of the game reduces to a clustering problem, where the increase in our bound is the distance function. This reduction leads to the first imperfect-recall abstraction algorithm with solution quality bounds.  We proceed to show a divide in the class of abstraction problems. If payoffs are at the same scale at all information sets considered for abstraction, the input forms a metric space, and this immediately yields a $2$-approximation algorithm for abstraction. Conversely, if this condition is not satisfied, we show that the input does not form a metric space.
  Finally, we provide computational experiments to evaluate the practical usefulness of the abstraction techniques. They show that running counterfactual regret minimization on such abstractions leads to good strategies in the original games.

%

\end{abstract}


\footnotetext[1]{This work was supported by the NSF under grants IIS-1320620, IIS-1546752, CCF-1101668, and IIS-0964579, the ARO under award W911NF-16-1-0061, as well as the CMU Center for Computational Thinking funded by Microsoft Research.}

\section{Introduction}

Game-theoretic equilibrium concepts provide a sound definition of how rational agents should act in multiagent settings.
%
To operationalize them, they have to be accompanied by techniques to compute equilibria, an important topic that has received significant attention in the literature
\cite{Lipton03:Playing,Littman03:Polynomial,Gilpin07:Lossless_long,Zinkevich07:Regret,Kroer15:Faster}.

Typically, equilibrium-finding algorithms do not scale to very large games.  This holds even for two-player zero-sum games (that can be solved in polynomial time~\cite{Koller96:Efficient}) when the games are large.  Therefore, the following has emerged as the leading framework for solving large extensive-form games~\cite{Sandholm10:State}.
First, the game is abstracted to generate a smaller game.  Then an ($\epsilon$-)Nash equilibrium is computed in the abstract game.  Then, the strategy from the abstract game is mapped back to the original game.
This framework, where \emph{imperfect-recall abstraction} is used to generate the smaller game~\cite{Waugh09:Practical}, is the leading approach in the Annual Computer Poker Competition (ACPC). For the last several years, the winning no-limit Texas Hold'em bots have employed it (for example, see the winners ``Baby Tartanian8'' by Brown and Sandholm and ``Slumbot'' by Jackson, from the ~\citet{poker_comp_website}).
Recently, this approach was also used in security games~\cite{Lisy16:Counterfactual}.
We expect that it could also be relevant to settings such as sequential auctions and trading, like the trading-agent competition, where abstraction approaches have previously been considered~\cite{Wellman05:Approximate}.
%
%
The motivation for imperfect-recall is computational. It allows the agent to forget less important details, in order to afford--from a computational perspective--to have a finer-grained abstraction of the present. \citet{Halpern13:Sequential} showed this same phenomenon in ``computational games,'' where agents are charged for computation. In this setting, choosing to forget information can be rational, because it decreases computational cost.

Ideally, abstraction would be performed in a lossless way, such that implementing an equilibrium from the abstract game results in an equilibrium in the full game. Lossless abstraction techniques were introduced by \citet{Gilpin07:Lossless_long} for a class of games called \textit{games of ordered signals}. Unfortunately, lossless abstraction often leads to games that are still too large to solve. Thus, we must turn to lossy abstraction.
However, significant abstraction \emph{pathologies} (\emph{nonmonotonicities}) have been shown in games while they cannot exist in single-agent settings: if an abstraction is refined, the equilibrium strategy from that new abstraction can actually be worse in the original game than the equilibrium strategy from a coarser abstraction~\cite{Waugh09:Abstraction}!
Until recently, all lossy abstraction techniques for general games of imperfect information were without any solution quality bounds. Then, \citet{Basilico11:Automated} gave bounds for the special game class called \textit{patrolling security game}.
\citet{Sandholm12:Lossy} provided lossy abstraction algorithms with bounds for stochastic games.
\citet{Lanctot12:no-regret} presented regret bounds in a class of imperfect-recall abstractions for equilibria computed using the \emph{counterfactual regret minimization algorithm (CFR)}~\cite{Zinkevich07:Regret}. \citet{Kroer14:Extensive-Form} showed solution quality bounds for a broad class of perfect-recall abstractions, and \citet{Kroer15:Discretization} extended these results to continuous action spaces. They left as an open problem whether similar bounds can be achieved for imperfect-recall abstractions, which are the state of the art in practical poker solving~\cite{Waugh09:Practical,Johanson13:Evaluating,Brown15:Hierarchical}.
\citet{Waugh15:Solving} and \citet{Brown15:Simultaneous}  develop iterative abstraction-refinement schemes that converge to a Nash equilibrium in the limit, but do not give bounds when an abstraction of the game is solved.

In contrast to prior work, our results are for a fairly general class of imperfect-recall abstractions, and are algorithm agnostic; they apply to both Nash equilibria, and strategies with bounded counterfactual regret.
We focus on these two classes of strategies because they are, to the best of our knowledge, the only types of strategies used in practical large-scale game solving. While refinements of Nash equilibria are desirable in some settings, they are usually too expensive to compute in the large-scale games where abstractions are applied. 
For solving large imperfect-information games---such as poker---an approximation of Nash equilibrium is usually computed (rather than a refinement of Nash equilibrium)~\cite{Sandholm10:State,Sandholm15:Solving}.  For example, that is what has been done for the leading program for two-player limit Texas Hold’em~\cite{Bowling15:Heads-up}  and for the leading programs for two-player no-limit Texas Hold’em~\cite{poker_comp_website,Brown15:Hierarchical,Jackson16:Compact}. Two-player limit Texas Hold'em has $10^{14}$ information sets in the game tree, while the no-limit variant has $10^{161}$~\cite{Johanson13:Measuring}.\footnote{Some authors have considered post-equilibrium computation techniques~\cite{Ganzfried15:Endgame}, but these are applied after the abstraction is both chosen and solved approximately. That said, we expect that our results extend to equilibrium refinements, and this would be an interesting extension.}

To develop our results, we generalize the notion of \emph{skew well-formed games} (SWFGs) introduced by \citet{Lanctot12:no-regret}, by introducing a new game class {\em \gameclassfull\ (\gameclass) games}. \gameclass\ games generalize SWFGs by also allowing chance error. Enabling chance error allows for a richer set of abstractions where nodes can go in the same abstract information set even if the nature probabilities of reaching those nodes and going from those nodes are not the same (this was left as an open problem by \citet{Lanctot12:no-regret}). This enables dramatically smaller abstractions for some games.  
We extend the techniques of \citet{Kroer14:Extensive-Form} to give theoretical solution-quality bounds for this class.
The solution quality bounds we derive are exponentially stronger than those of \citet{Lanctot12:no-regret} which had a linear dependence on the number of information sets in the game tree, and did not weight the leaf reward error by the probability of a given leaf being reached.
The reward error term in our result has only a linear dependence on tree height (actually, just the number of information sets any single player can experience on a path of play). Our leaf reward error term is weighted by the probability of the leaf being reached. Furthermore, our bounds are independent of the equilibrium computation method, while that prior work was only for CFR.


For games where abstraction of a subset of information sets at a single level is guaranteed to result in a \gameclass\ game, we show an equivalence between the problem of computing a single-level abstraction that  minimizes our theoretical solution-quality guarantees and a class of clustering problems. Using the decrease in solution-quality bound from abstracting a pair of information sets as a distance function, we show that such abstraction problems form a metric space. This yields a $2$-approximation algorithm for performing abstraction at a single level in the game tree when information sets differ only by the actions taken by players. When information sets differ based on nature's actions, our equivalence yields a new clustering objective that has not, to our knowledge, been previously studied. Our clustering results yield the first abstraction algorithm for computing imperfect-recall abstractions with solution-quality bounds. 

Finally, we use our theoretical results to conduct experiments on a simple die-based poker game that has been used as a benchmark for game abstraction in prior work. The experiments show that the CFR algorithm works well even on abstractions where different nature probabilities are abstracted together, and that the theoretical bound is within~$0$ to~$2$ orders of magnitude of the regrets at CFR convergence.


\section{Extensive-form games and abstraction concepts}
An {\em extensive-form game} (EFG) $\Gamma$ is a tuple $\langle N,A,S,Z,\mathcal{H},\sigma_0,u,\mathcal{I}\rangle$. $N$ is the set of players. $A$ is the set of all actions. $S$ is a set of nodes corresponding to sequences of actions. They describe a tree with root node $r\in S$. At each node $s$, some Player $i$ is active with actions $A_s$, and each branch at $s$ denotes a different choice in $A_s$. Let $\child{s}{a}$ be the node transitioned to by performing action $a\in A_s$ at node $s$. The set of all nodes where Player $i$ is active is called $S_i$.
The set of leaf nodes is denoted by $Z\subset S$. For each leaf node $z$, Player $i$ receives a reward of $u_i(z)$. We assume, WLOG., that all utilities are non-negative. $Z_s$ is the subset of leaf nodes reachable from a node $s$. 
$\mathcal{H}_i\subseteq \mathcal{H}$ is the set of heights in the game tree where Player $i$ acts. $\mathcal{H}_0$ is the set of heights where nature acts. $\sigma_0$ specifies the probability distribution for nature, with $\sigma_0(s,a)$ denoting the probability of nature choosing outcome $a$ at node $s$.
$\mathcal{I}_i\subseteq\mathcal{I}$ is the set of information sets where Player $i$ acts. $\mathcal{I}_i$ partitions $S_i$. For any two nodes $s_1,s_2\in I\in\mathcal{I}_i$, Player $i$ cannot distinguish among them, and $A_{s_1}=A_{s_2}$. We let $X(s)$ denote the set of information set and action pairs $I,a$ in the sequence leading to a node $s$, including nature. We let $X_{-i}(s),X_i(s)\subseteq X(s)$ be the subset of this sequence  such that actions by the subscripted player(s) are excluded or exclusively chosen.
We let $X^b(s)$ be the set of possible sequences of actions players can take in the subtree below $s$, with $X_{-i}^b(s),X_{i}^b(s)$ being the set of future sequences excluding or limited to Player $i$, respectively. We denote elements in these sets as $\vec{a}$.
$X^b(s,\vec{a}),X_{-i}^b(s,\vec{a}),X_{i}^b(s,\vec{a})$ are the analogous sets limited to sequences that are consistent with the sequence of actions $\vec{a}$. 
We let the set of leaf nodes reachable from $s$ for a particular sequence of actions $\vec{a} \in X^b(s)$ be $Z_{s}^{\vec{a}}$
For an information set $I$ on the path to a leaf node $z$, $z[I]$ denotes the predecessor $s\in I$ of $z$.

\emph{Perfect recall} means that no player forgets anything that the player observed in the past. Formally, for every Player $i\in N$, information set $I\in \mathcal{I}_i$, and nodes $s_1,s_2 \in I:X_i(s_1)=X_i(s_2)$. Otherwise, the game has {\em imperfect recall}. The most important consequence of imperfect recall is that a player can affect the distribution over nodes in their own information sets, as nodes in an information set may originate from different past information sets of the player.
For games $\Gamma^\prime =\langle N,A,S,Z,\mathcal{H},\sigma_0,u,\mathcal{I}^\prime\rangle$ and  $\Gamma =\langle N,A,S,Z,\mathcal{H},\sigma_0,u,\mathcal{I}\rangle$, we say that $\Gamma$ is a perfect-recall refinement of $\Gamma^\prime$ if $\Gamma$ has perfect-recall, and for any information set $I \in \mathcal{I}:\exists I^\prime\in\mathcal{I}^\prime, I \subseteq I^\prime$. That is, the game $\Gamma$ can be obtained by partitioning the nodes of each information set in $\mathcal{I}^\prime$ appropriately. For any perfect-recall refinement $\Gamma$, we let $\mathcal{P}(I^\prime)$ denote the information sets $I\in \mathcal{I}$ such that $I\subseteq I^\prime$ and $\bigcup_{I \in \mathcal{P}(I^\prime)} = I^\prime$. For an information set $I$ in $\Gamma$, we let $f_I$ denote the corresponding information set in $\Gamma^\prime$.

We will focus on the setting where we start out with some perfect-recall game $\Gamma$, and wish to compute an imperfect-recall abstraction such that the original game is a perfect-recall refinement of the abstraction. Imperfect-recall abstractions will be denoted by $\Gamma^\prime = \langle N,A,S,Z,\mathcal{H},\sigma_0,u,\mathcal{I^\prime}\rangle$. That is, they are the same game, except that some information sets have been merged.

We denote by $\sigma_i$ a {\em behavioral strategy} for Player $i$. For each information set $I$ where it is the player's turn to move, it assigns a probability distribution over $A_I$, the actions at the information set. $\sigma_i(I,a)$ is the probability of playing action $a$.
A {\em strategy profile} $\sigma=(\sigma_0,\ldots,\sigma_n)$ consists of a behavioral strategy for each player. We will often use $\sigma(I,a)$ to mean $\sigma_i(I,a)$, since the information set uniquely specifies which Player $i$ is active.
As described above, randomness external to the players is captured by the nature outcomes $\sigma_0$. We let $\sigma_{I\rightarrow a}$ denote the strategy profile obtained from $\sigma$ by having Player $i$ deviate to taking action $a$ at $I\in \mathcal{I}_i$.
Let the probability of going from node $s$ to a descendant $\hat{s}$ under strategy profile $\sigma$ be
  $\pi^\sigma(s, \hat s)=\Pi_{\tuple{\bar s, \bar a} \in X_{s, \hat s}} \sigma (\bar s, \bar a)$
where $X_{s, \hat s}$ is the nonempty set of pairs of nodes and actions on the path from $s$ to $\hat s$.
We let the probability of reaching node $s$ be $\pi^\sigma (s)=\pi^\sigma(r, s)$, the probability of going from the root node $r$ to $s$. Let $\pi^\sigma (I)=\sum_{s\in I} \pi^\sigma (s)$ be the probability of reaching any node in $I$. For probabilities over nature, $\pi^\sigma_0(s)=\pi^{\bar{\sigma}}_0(s)$ for all $\sigma,\bar\sigma,s\in S_0$, so we can ignore the superscript and write $\pi_0$.

For all definitions, the subscripts $i,-i$ refer to the same definition, but exclusively over or excluding Player $i$ in the product of probabilities, respectively.

%


For information set $I$ and action $a\in A_I$ at level $k\in \mathcal{H}_i$, we let $\mathcal{D}_{I}^{a}$ be the set of information sets at the next level in $\mathcal{H}_i$ reachable from $I$ when taking action $a$. Similarly, we let $\mathcal{D}_I^l$ be the set of descendant information sets at height $l\leq k$, where $\mathcal{D}_I^k=\{I\}$. Finally, we let $\mathcal{D}_s^{\vec{a},j}$ be the set of information sets reachable from node $s$ when action-vector $\vec{a}$ is played with probability one.

\subsection{Value functions}
We define value functions both for nodes and for
information sets.
  The value for Player $i$ of a given node $s$ under strategy profile
  $\sigma$ is $V_i^\sigma(s)=\sum_{z\in Z_s}\pi^\sigma(s, z) u_i(z)$.
  We use the definition of counterfactual value of an information set~\cite{Zinkevich07:Regret} to reason about the value of an information set.  The counterfactual value of an information set $I$ is the expected utility of the information set, assuming that all players follow strategy profile $\sigma$, except that Player $i$ plays to reach $I$. We normalize this by the probability of reaching the information set. 
  For a perfect-recall game $\Gamma$, the counterfactual value for Player $i$ of a given information set $I$ under strategy
  profile $\sigma$ is
\begin{align*}
  V_{i}^{\sigma}(I)=
  \begin{cases}
    \sum_{s\in I}\frac{\pi^\sigma_{-i}(s)}{\pi^\sigma_{-i}(I)}\sum_{z\in Z_{s}}\pi(s, z)u_{i}(z) & \mbox{if } \pi^\sigma_{-i}(I) > 0 \\
    0 & \mbox{if } \pi^\sigma_{-i}(I) = 0
  \end{cases}.
\end{align*}
\label{def:value-information-sets}
For the information set $I_r$ that contains just the root node $r$, we have $V^\sigma_i(I_r)=V^\sigma_i(r)$, which is the value of playing the game with strategy
profile $\sigma$. WLOG. we assume that at the root node it is not nature's turn to move. For imperfect-recall information sets, we let $W(I^\prime)=\sum_{s\in I^\prime} \frac{\pi^{\sigma}(s)}{\pi^{\sigma}(I^\prime)} V(s)$ be the value of an information set.

In perfect-recall games, for information set $I$ at height $k\in \mathcal{H}_i$, $V_{i}^{\sigma}(I)$ can be written as a sum over
descendant information sets at height $\hat{k}\in \mathcal{H}_i$, where $\hat{k}$ is the next level where Player $i$ acts~\cite{Kroer14:Extensive-Form}):
$
  V_{i}^{\sigma}(I)=\sum_{a\in A_{I}}\sigma(I,a)\sum_{\hat{I}\in \mathcal{D}_{I}^{a}}\frac{\pi^\sigma_{-i}(\hat{I})}{\pi^\sigma_{-i}(I)}V_{i}^{\sigma}(\hat{I}).
$
We will later be concerned with a notion of how much better a player $i$ could have done at an information set:
the regret for information set $I$ and action $a$ is $r(I,a)= V_i^{\sigma_{I\rightarrow a}} (I) - V_i^\sigma (I)$.
That is, the increase in expected utility for Player $i$ obtained by deviating to taking action $a$ at $I$. The immediate regret at an information set $I$ given a strategy profile $\sigma$ is $r(I)=\max_{a \in A_I} r(I,a)$. Regret is defined analogously for imperfect-recall games using $W(I)$.

\subsection{Equilibrium concepts}
In this section we define the equilibrium concepts we use.
We start with two classics.
\begin{definition}[$\epsilon$-Nash and Nash equilibria]
  An $\epsilon$-Nash equilibrium 
  is a strategy profile $\sigma$ such that for all $i$, $\bar \sigma_i$:
    \ $V_{i}^{\sigma}(r)+\epsilon \geq V_{i}^{\sigma_{-i}, \bar \sigma_i}(r)$. A Nash equilibrium is an $\epsilon$-Nash equilibrium where $\epsilon=0$.
\end{definition}
We will also use the concept of a \emph{self-trembling equilibrium}, introduced by \citet{Kroer14:Extensive-Form}.
It is a Nash equilibrium where the player assumes that opponents make no mistakes, but she might herself make mistakes, and thus her strategy must be optimal for all information sets that she could mistakenly reach by her own fault.
\begin{definition}[Self-trembling equilibrium]
  For a game $\Gamma$, a strategy profile $\sigma$ is a self-trembling equilibrium if it satisfies two conditions. First, it must be a Nash equilibrium. Second, for any information set $I\in \mathcal{I}_i$ such that $\pi^\sigma_{-i}(I)>0$, 
  and for all alternative strategies $\bar{\sigma}_{i}$,
$    V_{i}^{\sigma}(I)\geq V_{i}^{\sigma_{-i},\bar{\sigma}_{i}}(I)$.
  We call this second condition the {\em self-trembling} property.
  \label{def:self_trembling_equilibrium}
\end{definition}
An $\epsilon$-self-trembling equilibrium is defined analogously, for each information set $I\in \mathcal{I}_i$, we require $V_{i}^{\sigma}(I)\geq V_{i}^{\sigma_{-i},\bar{\sigma}_{i}}(I)-\epsilon$. For imperfect-recall games, the property $\pi^\sigma_{-i}(I^\prime)>0$ does not give a probability distribution over the nodes in an information set $I^\prime$, since Player $i$ can affect the distribution over the nodes. For such information sets, it will be sufficient for our purposes to assume that $\sigma_i$ is (approximately) utility maximizing for some (arbitrary) distribution over $\mathcal{P}(I^\prime)$: our bounds are the same for any such distribution.

\subsection{\Gameclassfull\ (\gameclass) games}
Now we introduce the class of imperfect-recall games that we consider as potential abstractions of a perfect-recall game. We call this class \gameclass\ games. A \gameclass\ game is an imperfect-recall game where there exists a perfect-recall refinement of the game that satisfies a certain set of properties that we introduce below. We will focus on the general problem of computing solution concepts in \gameclass\ games and mapping the solution concept to a perfect-recall refinement. Typically, the perfect-recall refinement is the original game, and the \gameclass\ game is an abstraction that is easier to solve.

Intuitively, a \gameclass\ game is an imperfect-recall game where there exists a refinement that satisfies two intuitive properties for any pair of information sets that are separated in the perfect-recall refinement. The first is that a bijection exists between the leaf nodes of the information set pair such that leaves mapped to each other pass through the same non-nature actions on the path from the information set to the leaf. This ensures that the probability of reaching pairs of leaves that map to each other is similar. The second is that a bijection exists between the nodes in the pair of information sets, such that the path leading to two nodes mapped to each other passes through the same information set-action pairs over all players except the acting player and nature. This ensures that the conditional distribution over nodes in the information sets is similar.
\begin{definition}
  For an EFG $\Gamma^\prime$, and a perfect-recall refinement $\Gamma$, we say that $\Gamma^\prime$ is a {\em \gameclass\ game with respect to $\Gamma$} if for all $i\in N, I^\prime\in \mathcal{I}_i^\prime, I,\breve{I}\in \mathcal{P}(I^\prime)$, there exists a bijection $\phi:Z_{I} \rightarrow Z_{\breve{I}}$ such that for all $z \in Z_{I}$:
  \begin{enumerate}
  \item In $\Gamma^\prime$, $X_{-\{i,0\}}(z)=X_{-\{i,0\}}(\phi(z))$, \label{con:opponent-information-structure} that is, for two leaf nodes mapped to each other (for these two information sets in the original game), the action sequences of the other players on those two paths must be the same in the abstraction.\footnote{It is possible to relax this notion  slightly: if two actions of another player are not the same, as long as they are on the path (at the same level) to all nodes in their respective full-game information sets ($I$ and $\breve{I}$), they do not affect the distribution over nodes in the information sets, and are thus allowed to differ in the abstraction.} 
      \item In $\Gamma^\prime$, $X_{i}(z[I],z)=X_{i}(\phi(z)[\breve{I}],\phi(z))$, \label{con:player-information-structure} that is, for two leaf nodes mapped to each other (for information sets $I$ and $\breve{I}$ in the original game), the action sequence of Player $i$ from $I$ to $z$ and from $\breve{I}$ to $\phi(z)$ must be the same in the abstraction.
  \end{enumerate}
  \label{def:gameclass}
\end{definition}
Our definition implicitly assumes that leaf nodes are all at the same level. This is without loss of generality, as any perfect-recall game can be extended to satisfy this.

With this definition, we can define the following error terms for a \gameclass\ refinement $\Gamma$ of an imperfect-recall game $\Gamma'$ for all $i\in N, I^\prime\in \mathcal{I}_i^\prime, I,\breve{I}\in \mathcal{P}(I^\prime),z\in I$ 
\begin{itemize}
\item $\left| u_i(z) - \delta_{I, \breve{I}} u_i(\phi (z)) \right| \leq \epsilon^R_{I, \breve{I}}(z) $\label{con:reward-similar}, the reward error at $z$, after scaling by $\delta_{I, \breve{I}}$ at  $\breve{I}$. 
  \vspace{-1.5mm}
\item $\left| \pi_0(z[I], z) - \pi_0(\phi(z)[\breve{I}], \phi(z)) \right| = \epsilon^0_{I, \breve{I}}(z)$ \label{con:leaf-probability-similar}, the leaf probability error at $z$. 
  \vspace{-1.5mm}
  \item $\left| \frac{\pi_0(z[I])}{\pi_0(I)} - \frac{\pi_0(\phi(z)[\breve{I}])}{\pi_0(\breve{I})}\right| = \epsilon^D_{I,\breve{I}}(z[I])$ \label{con:distribution-similar}, the distribution error of $z[I]$. 
\end{itemize}
The scaling term $\delta_{I, \breve{I}}$ does not affect the abstract game; it can be chosen by the user to minimize the bounds proved later. The reward error uses an inequality rather than equality because the error at a leaf node can vary by player.

Instead of our probability and distribution error terms, \citet{Lanctot12:no-regret} require $\pi_0(z) = l_{I, \breve{I}} \pi_0(\phi_{I,\breve{I}}(z))$, where $l_{I, \breve{I}}$ is a scalar defined on a per information set pair basis. We omit any such constraint, and instead introduce probability and distribution error terms as above. Our definition allows for a richer class of abstractions. Consider some game where every nature probability in the game differs by a small amount. For such a game, no two information sets can be merged according to the SWFG definition, whereas our definition allows such abstraction. 

We define $\overline{u}_{I,\breve{I}}(s)=\max_{i\in N,z\in Z_s}  u_i(z) + \epsilon_{I,\breve{I}}^R(z)$, the maximum utility plus its scaled error achieved at any leaf node. This will simplify notation when we take the maximum over error terms related to probability transitions. 

We now define additional aggregate approximation error terms. These will be useful when reasoning inductively about more than one height of the game at a time. 
We define the {\em reward  approximation error} $  \epsilon^{R}_{I,\breve{I}}(s)$ for information sets $I,\breve{I}\in \mathcal{P}(I^\prime)$ and any node $s$ in $\breve{I}$ to be
\begin{align*}
  \epsilon^{R}_{I,\breve{I}}(s) = \begin{cases}
    \epsilon^{R}_{I,\breve{I}}(z)  & \mbox{if } \exists z\in Z:z = s\\
    \sum_{a\in A_s} \sigma_0(s,a) \epsilon^{R}_{I,\breve{I}}(\child{s}{a}) & \mbox{if } s \in S_{0}  \\
    \max_{a\in A_s}  \epsilon^R_{I,\breve{I}}(\child{s}{a}) & \mbox{if }  s \notin S_{0} \land s \notin Z  \\
  \end{cases},\quad
\end{align*}
We define the {\em transition approximation error} $\epsilon^{0}_{I,\breve{I}}(s) $ for information sets $I,\breve{I}\in \mathcal{P}(I^\prime)$ and any node $s$ in $\breve{I}$ to be
\begin{align*}
  \epsilon^{0}_{I,\breve{I}}(s) = \max_{\vec{a} \in X^b_{-0}(s)}\sum_{z \in Z_{s}^{\vec{a}}}\epsilon^0_{I, \breve{I}}(z)\overline{u}_{I,\breve{I}}(s)
\end{align*}
We define the {\em distribution approximation error} for an information set pair $I,\breve{I} \in \mathcal{P}(I^\prime)$ to be
\begin{align*}
  \epsilon^D_{I,\breve{I}} = \sum_{s \in I} \epsilon^D_{I,\breve{I}}(s) \overline{u}_{I,\breve{I}}(s).
\vspace{-5mm}
\end{align*}

\begin{figure}[]
  \scalebox{0.61}{
    \definecolor{c1}{RGB}{27,158,119}
\definecolor{c2}{RGB}{217,95,2}
\definecolor{c3}{RGB}{117,112,179}
\definecolor{c4}{RGB}{231,41,138}
\definecolor{c5}{RGB}{102,166,30}
\definecolor{c6}{RGB}{230,171,2}
\forestset{
leftedge/.style={
  edge label={node[inner sep=1pt,midway,auto,swap,font=\scriptsize]{$#1$}}
  },
rightedge/.style={
  edge label={node[inner sep=1pt,midway,auto,font=\scriptsize]{$#1$}}
  },
leftedgeb/.style={
  edge label={node[inner sep=1pt,near end,auto,swap,font=\scriptsize]{$#1$}}
  },
rightedgeb/.style={
  edge label={node[inner sep=1pt,near end,auto,font=\scriptsize]{$#1$}}
  }
}
\begin{forest}
for tree={
  parent anchor=south,
  child anchor=north,
  math content,
  text height=2ex,
  l sep=25pt,
  s sep=20pt,
  where level={3}{s sep=30pt}{}
}
[,phantom,
  [P2,circle,draw,name=p2n1
    [N,circle,draw,leftedge={a}
      [P1,circle,draw,name=n1,leftedge={$0.5$}
        [$10$,leftedgeb={l}
        ]
        [$0$,rightedgeb={r}
        ]
      ]
      [P1,circle,draw,name=n2,rightedge={$0.5$}
        [$0$,leftedgeb={l}
        ]
        [10,rightedgeb={r}
        ]
      ]
    ]
    [N,circle,draw,rightedge={b}
      [P1,circle,draw,name=n3,leftedge={$0.5$}
        [$10$,leftedgeb={L}
        ]
        [$0$,rightedgeb={R}
        ]
      ]
      [P1,circle,draw,name=n4,rightedge={$0.5$}
        [$0$,leftedgeb={L}
        ]
        [$9$,rightedgeb={R}
        ]
      ]
    ]
  ]
  [P2,circle,draw,name=p2n2
    [N,circle,draw,leftedge={A}
      [P1,circle,draw,name=n5,leftedge={$0.4$}
        [$10$,leftedgeb={l}
        ]
        [$0$,rightedgeb={r}
        ]
      ]
      [P1,circle,draw,name=n6,rightedge={$0.6$}
        [$0$,leftedgeb={l}
        ]
        [$10$,rightedgeb={r}
        ]
      ]
    ]
    [N,circle,draw,rightedge={B}
      [P1,circle,draw,name=n7,leftedge={$0.5$}
        [$10$,leftedgeb={L}
        ]
        [$0$,rightedgeb={R}
        ]
      ]
      [P1,circle,draw,name=n8,rightedge={$0.5$}
        [$0$,leftedgeb={L}
        ]
        [$10$,rightedgeb={R}
        ]
      ]
    ]
  ]
]
\draw[dashed] (n1) to[out=east,in=west] (n2);
\draw[dashed] (n3) to[out=east,in=west] (n4);
\draw[dashed] (n5) to[out=east,in=west] (n6);
\draw[dashed] (n7) to[out=east,in=west] (n8);
\draw[dashed,<->,color=c3,line width = 1.2pt] (n2) to[out=12,in=168] (n6);
\draw[dashed,<->,color=c3,line width = 1.2pt] (n1) to[out=12,in=168] (n5);
\draw[dashed,<->,color=c4,line width = 1.2pt] (n3) to[out=-12,in=-168] (n7);
\draw[dashed,<->,color=c4,line width = 1.2pt] (n4) to[out=-12,in=-168] (n8);
\draw[dashed,<->,color=c2,line width = 1.2pt] (p2n1) to[out=east,in=west] (p2n2);
\end{forest}
  }
  \caption{Two subtrees of a game tree. Information sets are denoted by dotted lines. A \gameclass\ abstraction is shown, with merged information sets and their node mapping denoted by dotted lines with arrowheads. All actions are mapped to their corresponding upper/lower-case actions in the merged information sets.}
  \label{fig:error_prop}
\end{figure}
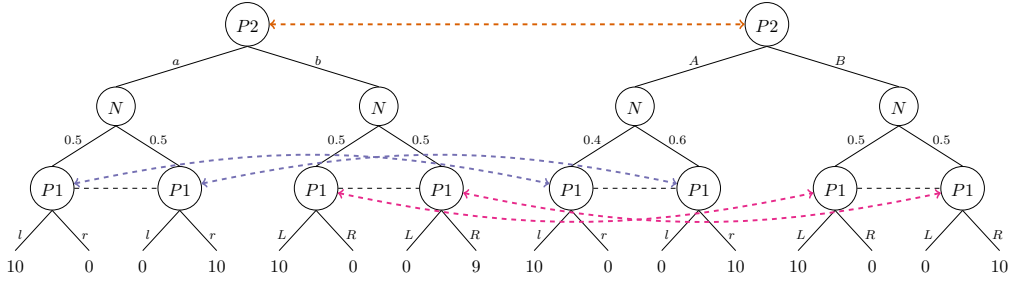

Figure~\ref{fig:error_prop} shows two subtrees of an example game tree. Dotted lines with arrow heads show a \gameclass\ abstraction of the game. First consider the left node for P2, which maps to the right P2 node. It has distribution approximation error of zero (as is always the case for singleton information sets). It has transition approximation error of $0.2\cdot 10=2$ (the maximizing sequences of player actions $\vec{a}$ are $[a,l]$ or  $[a,r]$). Finally, the node has reward approximation error $1\cdot 0.5$, since the biggest utility difference between nodes mapped to each other is $1$, and the definition of reward approximation error allows taking a weighted sum at nature nodes. Now consider the leftmost information set for P1. The distribution approximation error at this information set is $0.2\cdot 10$, since the conditional probability of being at each of the two nodes in the information set differs by $0.1$ from the node in the information set that it is mapped to. The transition approximation error is zero for both nodes in the information set. The reward approximation error is zero, since all leaf nodes under the information set are mapped to leaf nodes with the same utility.

\section{Strategies from abstract near-equilibria have bounded regret} \label{sec:bounded-regret-equilibria}
To prove our main result, we first show that strategies with bounded regret at information sets in \gameclass\ games have bounded regret at their perfect-recall refinements. 
\begin{proposition}
For any \gameclass\ game $\Gamma^{\prime}$, refinement $\Gamma$,
strategy profile $\sigma$, and information set $I^{\prime}\in\mathcal{I}^{\prime}$ such that Player $i$ has bounded regret $r(I^\prime,a)$ for all $a\in A_{I^\prime}$,
the regret $r(I,a^{*})$ at any information set $I\in\mathcal{P}(I^{\prime})$
and action $a^{*}\in A_{I}$ is bounded by
\begin{align*}
  r(I,a^{*}) \leq & \max_{\breve{I}\in\mathcal{P}(I^{\prime})}\delta_{I,\breve{I}}r(I^{\prime},a^{*}) + 2 \sum_{s\in I}\frac{\pi^{\sigma}(s)}{\pi^{\sigma}(I)}  \left( \epsilon_{I,\breve{I}}^{0}(s) + \epsilon_{I,\breve{I}}^{R}(s) \right) + \epsilon_{I,\breve{I}}^{D}.
\end{align*}
\label{pro:bounded-regret-action}
\vspace{-2mm}
\end{proposition}
\begin{proof}
Given some $I^\prime$ such that $\pi^{\sigma}_{-i}(I^\prime)>0$, we assume that $\pi^{\sigma}_i(I^\prime)>0$. For information sets where this is not the case, we assume any distribution over the choices of Player $i$  leading to $I^\prime$. Note that other players cannot affect the distribution over $\mathcal{P}(I^\prime)$ due to Condition~\ref{con:opponent-information-structure} of Definition~\ref{def:gameclass}. By the definition of regret of an action, we have:
\begin{align*}
& r(I^{\prime},a^{*})= W^{\sigma_{I\rightarrow a^{*}}}(I^{\prime})-W^{\sigma}(I^{\prime})\\
= & \sum_{s^{\prime}\in I^{\prime}}\frac{\pi^\sigma(s^{\prime})}{\pi^\sigma(I^{\prime})}\sum_{z^{\prime}\in Z_{\child{s^\prime}{a^*}}}\pi^\sigma(\child{s^\prime}{a^*} z^{\prime})u_{i}(z^{\prime}) - \sum_{s^{\prime}\in I^{\prime}}\frac{\pi^\sigma(s^{\prime})}{\pi^\sigma(I^{\prime})}\sum_{a\in A_{I}}\pi^\sigma(I^{\prime},a)\sum_{z^{\prime}\in Z_{\child{s^\prime}{a}}}\pi^\sigma(\child{s^\prime}{a}, z^{\prime})u_{i}(z^{\prime})\\
= & \sum_{\breve{I}\in\mathcal{P}(I^{\prime})}\sum_{s^{\prime}\in\breve{I}}\frac{\pi^\sigma(s^{\prime})}{\pi^\sigma(I^{\prime})} \left(\sum_{z^{\prime}\in Z_{\child{s^\prime}{a^*}}}\pi^\sigma(\child{s^\prime}{a^*}, z^{\prime})u_{i}(z^{\prime}) - \sum_{a\in A_{I}}\pi^\sigma(I^{\prime},a)\sum_{z^{\prime}\in Z_{\child{s^\prime}{a}}}\pi^\sigma(\child{s^\prime}{a}, z^{\prime})u_{i}(z^{\prime}) \right)
\end{align*}

Note that $\sum_{\breve{I}\in\mathcal{P}(I^{\prime})}\sum_{s^{\prime}\in\breve{I}}\frac{\pi^\sigma(s^{\prime})}{\pi^\sigma(I^{\prime})}=\sum_{\breve{I}\in\mathcal{P}(I')}\frac{\pi^\sigma(\breve{I})}{\pi^\sigma(I^{\prime})}$ sums over a probability distribution on $\mathcal{P}(I^{\prime})$.
We take the minimum over this distribution:
\begin{align*}
\geq & \min_{\breve{I}\in\mathcal{P}(I^{\prime})}\sum_{s^{\prime}\in\breve{I}}\frac{\pi^\sigma(s^{\prime})}{\pi^\sigma(\breve{I})}  \left(\sum_{z^{\prime}\in Z_{\child{s^\prime}{a^*}}}\pi^\sigma(\child{s^\prime}{a^*} , z^{\prime})u_{i}(z^{\prime}) - \sum_{a\in A_{I}}\pi^\sigma(I^{\prime},a)\sum_{z^{\prime}\in Z_{\child{s^\prime}{a}}}\pi^\sigma(\child{s^\prime}{a} , z^{\prime})u_{i}(z^{\prime})\right)
\end{align*}

Let $I_{m}$ 
be the minimizer of the expression above. Now we can bound the value using the reward approximation error term:
\begin{align*}
= & \sum_{s^{\prime}\in I_{m}}\frac{\pi^\sigma(s^{\prime})}{\pi^\sigma(I_{m})} \left(\sum_{z^{\prime}\in Z_{\child{s^\prime}{a^*}}}\pi^\sigma (\child{s^\prime}{a^*} , z^{\prime})u_{i}(z^{\prime})-\sum_{a\in A_{I}}\pi^\sigma(I^{\prime},a)\sum_{z^{\prime}\in Z_{\child{s^\prime}{a}}}\pi^\sigma(\child{s^\prime}{a} , z^{\prime})u_{i}(z^{\prime})\right)\\
\geq & \frac{1}{\delta_{I,I_{m}}} \left(\sum_{s^{\prime}\in I_{m}}\frac{\pi^\sigma(s^{\prime})}{\pi^\sigma(I_{m})} \left(\sum_{z^{\prime}\in Z_{\child{s^\prime}{a^*}}}\pi^\sigma(\child{s^\prime}{a^*} , z^{\prime})u_{i}(\phi_{I_{m},I}(z^{\prime})) \right.\right.\\
&\left.\left. -\sum_{a\in A_{I}}\pi^\sigma(I^{\prime},a)\sum_{z^{\prime}\in Z_{\child{s^\prime}{a}}}\pi^\sigma(\child{s^\prime}{a} , z^{\prime})u_{i}(\phi_{I_{m},I}(z^{\prime}))\right)-\frac{1}{\delta_{I,I_{m}}} 2\epsilon_{I,I_{m}}^{R}(s^\prime) \right)
\end{align*}
Multiplying both sides by $\delta_{I,I_{m}}$ gives
\begin{align*}
\delta_{I,I_{m}} r(I^{\prime},a^{*}) \geq & \sum_{s^{\prime}\in I_{m}}\frac{\pi^\sigma(s^{\prime})}{\pi^\sigma(I_{m})} \left(\sum_{z^{\prime}\in Z_{\child{s^\prime}{a^*}}}\pi^\sigma(\child{s^\prime}{a^*} , z^{\prime})u_{i}(\phi_{I_{m},I}(z^{\prime})) \right.\\
& \left. -\sum_{a\in A_{I}}\pi^\sigma(I^{\prime},a)\sum_{z^{\prime}\in Z_{\child{s^\prime}{a}}}\pi^\sigma(\child{s^\prime}{a} , z^{\prime})u_{i}(\phi_{I_{m},I}(z^{\prime})) - 2\epsilon_{I,I_{m}}^{R}(s^\prime)\right)
\end{align*}
To ease notation, let $s^\prime=z^{\prime}[I_m]$, $s=\phi_{I_m,I}(z^\prime)[I]$, and similarly $z=\phi_{I_m,I}(z^\prime)$.
Now we can apply the distribution approximation error:
\begin{align*}
\geq & \sum_{s^{\prime}\in I_{m}} \left(\frac{\pi^{\sigma}(s)}{\pi^{\sigma}(I)}-\epsilon_{I,I_{m}}^{D}(s)\right)  \left(\sum_{z^{\prime}\in Z_{\child{s^\prime}{a^*}}}\pi^\sigma(\child{s^\prime}{a^*} , z^{\prime})u_{i}(\phi_{I_{m},I}(z^{\prime})) \right.\\
& \left. -\sum_{a\in A_{I}}\pi^\sigma(I^{\prime},a)\sum_{z^{\prime}\in Z_{\child{s^\prime}{a}}}\pi^\sigma(\child{s^\prime}{a} , z^{\prime})u_{i}(\phi_{I_{m},I}(z^{\prime})) - 2\epsilon_{I,I_{m}}^{R}(s^\prime)\right)
\end{align*}

 For all $a\in A_I$, $\sum_{s^\prime\in \breve{I}}  \sum_{z^{\prime}\in Z_{\child{s^\prime}{a}}}$ can be rewritten as the sum $\sum_{s \in I}  \sum_{z \in Z_{\child{s}{a}}}$ as Condition~\ref{con:player-information-structure} of~Definition~\ref{def:gameclass} ensures that if $(\breve{I},a)$ is on the path to $z$, then $(I,a)$ is on the path to $\phi_{I_m,I}(z)$. This gives us
\begin{align*}
  = & \sum_{s\in I} \left(\frac{\pi^{\sigma}(s)}{\pi^{\sigma}(I)}-\epsilon_{I,I_{m}}^{D}(s)\right)  \left( \sum_{z\in Z_{\child{s}{a^*}}} \pi^\sigma_0(\child{s^\prime}{a^*} , z^{\prime})u_{i}(z) \right.\\
  & \left. -   \sum_{a\in A_{I}}\pi^\sigma(I^{\prime},a)\sum_{z\in Z_{\child{s}{a}}} \pi^\sigma_0(\child{s^\prime}{a} , z^{\prime})u_{i}(z)-2\epsilon_{I,I_{m}}^{R}(s) \right) \\
  \geq & \sum_{s\in I} \frac{\pi^{\sigma}(s)}{\pi^{\sigma}(I)}  \left( \sum_{z\in Z_{\child{s}{a^*}}} \pi^\sigma_0(\child{s^\prime}{a^*} , z^{\prime})u_{i}(z) \right.\\
  & \left. -   \sum_{a\in A_{I}}\pi^\sigma(I^{\prime},a)\sum_{z\in Z_{\child{s}{a}}} \pi^\sigma_0(\child{s^\prime}{a} , z^{\prime})u_{i}(z)-2\epsilon_{I,I_{m}}^{R}(s) \right) -\epsilon_{I,I_{m}}^{D}
\end{align*}

We rewrite the summation over  $Z_{\child{s}{a}}$ so that we first sum over the possible sequences of actions  $X^b_{-0}(\child{s}{a})=X^b_{-0}(\child{s^\prime}{a})$ players excluding nature can take after $\child{s}{a}$. We then sum over the possible sequences of actions nature can take for the chosen sequence $\vec{a}$. Since this uniquely specifies leaf nodes, we can treat elements of this summation as such. For any such node $s$ and leaf node $z$, $\pi^\sigma(s, z)=\pi^\sigma_{-0}(\vec{a})\pi_0(s,z)$. We use this observation along with the transition approximation error to get 
\begin{align*}
  \geq{} &\sum_{s\in I} \frac{\pi^{\sigma}(s)}{\pi^{\sigma}(I)} \left( \sum_{\vec{a}^* \in X^b_{-0}(\child{s}{a^*})} \pi^\sigma(\vec{a}^*)  \sum_{z \in Z_{s}^{\vec{a}^*}}  \pi^\sigma_0(\child{s}{a^*} , z) u_{i}(z) \right.\\
  & \left. - \sum_{\vec{a} \in X^b_{-0}(s)} \pi^\sigma(\vec{a}) \sum_{z \in Z_{s}^{\vec{a}}}  \pi^\sigma_0(\child{s}{a} , z) u_{i}(z) - 2\epsilon^0_{I,I_m}(s) -2\epsilon_{I,I_{m}}^{R}(s) \right) -\epsilon_{I,I_{m}}^{D}
\end{align*}
Rearranging terms gives us that this is exactly equal to
\begin{align*}
 & = V^{\sigma_{I\rightarrow a^{*}}}(I)-V^{\sigma}(I) - 2 \sum_{s\in I}\frac{\pi^{\sigma}(s)}{\pi^{\sigma}(I)}  \left( \epsilon_{I,I_{m}}^{0}(s) + \epsilon_{I,I_{m}}^{R}(s) \right) -\epsilon_{I,I_{m}}^{D}\\
 & = r(I,a^{*})-2 \sum_{s\in I}\frac{\pi^{\sigma}(s)}{\pi^{\sigma}(I)}  \left( \epsilon_{I,I_{m}}^{0}(s) + \epsilon_{I,I_{m}}^{R}(s) \right) -\epsilon_{I,I_{m}}^{D}
\end{align*}

Summarizing, this gives us
\begin{align*}
\delta_{I,I_{m}} r(I^{\prime},a^{*}) & \geq  r(I,a^{*})- 2 \sum_{s\in I}\frac{\pi^{\sigma}(s)}{\pi^{\sigma}(I)}  \left( \epsilon_{I,I_{m}}^{0}(s) + \epsilon_{I,I_{m}}^{R}(s) \right) -\epsilon_{I,I_{m}}^{D}\\
\Leftrightarrow r(I,a^{*})& \leq  \delta_{I,I_{m}}r(I^{\prime},a^{*}) + 2 \sum_{s\in I} \frac{\pi^{\sigma}(s)}{\pi^{\sigma}(I)}  \left( \epsilon_{I,I_{m}}^{0}(s) + \epsilon_{I,I_{m}}^{R}(s) \right) + \epsilon_{I,I_{m}}^{D}\\
& \leq  \max_{\breve{I}\in\mathcal{P}(I^{\prime})}\delta_{I,\breve{I}}r(I^{\prime},a^{*}) + 2 \sum_{s\in I} \frac{\pi^{\sigma}(s)}{\pi^{\sigma}(I)}  \left( \epsilon_{I,\breve{I}}^{0}(s) + \epsilon_{I,\breve{I}}^{R}(s) \right) + \epsilon_{I,\breve{I}}^{D}
\end{align*}
which completes the proof.
\end{proof}

 Intuitively, the scaling variable $\delta_{I,\breve{I}}$ ensures
that if the regret at $I^{\prime}$ is largely based
on some other information set, then the regret is scaled to fit with the payoffs at $I$.

With this result, we are ready to prove our main results. 
First, we show that strategies with bounded regret at each information set in \gameclass\ games are $\epsilon$-self-trembling equilibria when  implemented in any perfect-recall refinement.
\begin{theorem}
For any \gameclass\ game $\Gamma^{\prime}$ and strategy $\sigma$
with bounded immediate regret $r_{I^{\prime}}$ at each information
set $I^{\prime}\in\Gamma^{\prime}$ where $\sigma_{-i}(I^{\prime})>0$,
$\sigma$ is an $\epsilon$-self-trembling equilibrium when implemented
in any perfect-recall refinement $\Gamma$, where $\epsilon=\max_{i\in N}\epsilon_{i}$
and
\begin{align*}
  \epsilon_{i}= &
                  \max_{\vec{a} \in X_{i}^b(r)}   \sum_{j\in\mathcal{H}_{i}} \sum_{I \in \mathcal{D}_r^{\vec{a},j}} \pi_{-i}^\sigma(I)  \left(  \max_{\breve{I}\in\mathcal{P}(f_{I})} \delta_{I,\breve{I}}r(f_{I})                   + 2 \sum_{s\in I} \frac{\pi^{\sigma}(s)}{\pi^{\sigma}(I)}   \left( \epsilon_{I,\breve{I}}^{0}(s) + \epsilon_{I,\breve{I}}^{R}(s) \right) + \epsilon_{I,\breve{I}}^{D} \right). 
\end{align*}
\label{the:refinement-regret-bound-1}
\vspace{-2.8mm}
\end{theorem}
\begin{proof}
Consider some alternative strategy $\sigma^{*}$ where Player $i$
deviates to a best response and $\sigma_{-i}=\sigma_{-i}^{*}$. We
prove the bound by induction over the levels $\mathcal{H}_{i}$ belonging
to Player $i$. For the base case, consider any abstract information
set $I^{\prime}\in\mathcal{I}_{i}^{\prime}$ and any $I\in\mathcal{P}(I^{\prime})$
at the lowest level $l$ in $\mathcal{H}_{i}$. We know that no mixed strategy is better than the single best action when the strategies of the other players are held constant. This fact and Proposition \ref{pro:bounded-regret-action}
gives us that:
\begin{align*}
V^{\sigma^*}(I) &\leq \max_{a\in A_{I}}V^{\sigma_{I\rightarrow a}}(I) \\ 
&\leq V^{\sigma}(I) + \max_{\breve{I}\in\mathcal{P}(I^{\prime})}\delta_{I,\breve{I}}r(I^{\prime}) + 2 \sum_{s\in I}\frac{\pi^{\sigma}(s)}{\pi^{\sigma}(I)}   \left( \epsilon_{I,\breve{I}}^{0}(s) + \epsilon_{I,\breve{I}}^{R}(s) \right) + \epsilon_{I,\breve{I}}^{D}
\end{align*}
For the inductive step, we assume the following holds for all information sets $I$ at heights $l<k\in\mathcal{H}_{i}$:
\begin{align}
  V^{\sigma^*}(I) & \leq  V^{\sigma}(I)+\max_{\vec{a} \in X_{i}^b(I)}   \sum_{j\in\mathcal{H}_{i},j \leq l} \sum_{\hat{I} \in \mathcal{D}_I^{\vec{a},j}} \frac{\pi_{-i}^\sigma(\hat{I})}{\pi_{-i}^\sigma(I)} \psi(\hat{I}) \nonumber\\
  \psi(\hat{I}) & = \left(  \max_{\breve{I}\in\mathcal{P}(f_{\hat{I}})}\delta_{\hat{I},\breve{I}}r(f_{\hat{I}}) + 2 \sum_{s\in \hat{I}} \frac{\pi^{\sigma}(s)}{\pi^{\sigma}(\hat{I})}   \left( \epsilon_{\hat{I},\breve{I}}^{0}(s) + \epsilon_{\hat{I},\breve{I}}^{R}(s) \right) + \epsilon_{\hat{I},\breve{I}}^{D} \right) \label{equ:psi}
\end{align}
Now consider some information set $I$ at height $k$. We apply the inductive assumption to the value of an information set:
\begin{align*}
V^{\sigma^{*}}(I) & =  \sum_{a\in A_{I}}\sigma^{*}(I,a)\sum_{\hat{I}\in\mathcal{D}_{I}^{a}} \frac{\pi^{\sigma}_{-i}(\hat{I})}{\pi^{\sigma}_{-i}(I)} V^{\sigma^{*}}(\hat{I})\\
& \leq \sum_{a\in A_{I}}\sigma^{*}(I,a) \sum_{\hat{I}\in\mathcal{D}_{I}^{a}} \frac{\pi^{\sigma}_{-i}(\hat{I})}{\pi^{\sigma}_{-i}(I)} \left(V^{\sigma}(\hat{I})  +\max_{\vec{a} \in X_{i}^b(\hat{I})}   \sum_{j\in\mathcal{H}_{i},j < k} \sum_{\grave{I} \in \mathcal{D}_{\hat{I}}^{\vec{a},j}} \frac{\pi_{-i}^\sigma(\grave{I})}{\pi_{-i}^\sigma(\hat{I})} \psi(\grave{I}) \right)\\
& \leq  \max_{a\in A_{I}} \sum_{\hat{I}\in\mathcal{D}_{I}^{a}} \frac{\pi^{\sigma}_{-i}(\hat{I})}{\pi^{\sigma}_{-i}(I)} V^{\sigma}(\hat{I})  +\max_{\vec{a} \in X_{i}^b(I)}   \sum_{j\in\mathcal{H}_{i},j < k} \sum_{\grave{I} \in \mathcal{D}_{I}^{\vec{a},j}} \frac{\pi_{-i}^\sigma(\grave{I})}{\pi_{-i}^\sigma(I)} \psi(\grave{I}) \\
\vspace{-3mm}
\end{align*}
The last inequality is obtained by taking the maximum over $A_I$, splitting the terms,  and multiplying in $\frac{\pi_{-i}^\sigma(\hat{I})}{\pi_{-i}^\sigma(I)}$.
Now we can apply Proposition \ref{pro:bounded-regret-action}
to bound the immediate regret:
\begin{align*}
\vspace{-3mm}
  \leq {}& V^{\sigma}(I) + \max_{\breve{I}\in\mathcal{P}(f_I)}\delta_{I,\breve{I}}r(I^{\prime},a^{*}) + 2 \sum_{s\in I} \frac{\pi^{\sigma}(s)}{\pi^{\sigma}(I)}  \left( \epsilon_{I,\breve{I}}^{0}(s) + \epsilon_{I,\breve{I}}^{R}(s) \right) + \epsilon_{I,\breve{I}}^{D} \\
  & +\max_{\vec{a} \in X_{i}^b(I)}   \sum_{j\in\mathcal{H}_{i},j < k} \sum_{\grave{I} \in \mathcal{D}_{I}^{\vec{a},j}} \frac{\pi_{-i}^\sigma(\grave{I})}{\pi_{-i}^\sigma(I)} \psi(\grave{I}) \\
  = {}& V^{\sigma}(I) + \psi(I) +\max_{\vec{a} \in X_{i}^b(I)}   \sum_{j\in\mathcal{H}_{i},j < k} \sum_{\grave{I} \in \mathcal{D}_{I}^{\vec{a},j}} \frac{\pi_{-i}^\sigma(\grave{I})}{\pi_{-i}^\sigma(I)} \psi(\grave{I}) \\
  = {}& V^{\sigma}(I)  +\max_{\vec{a} \in X_{i}^b(I)}   \sum_{j\in\mathcal{H}_{i},j \leq k} \sum_{\grave{I} \in \mathcal{D}_{I}^{\vec{a},j}} \frac{\pi_{-i}^\sigma(\grave{I})}{\pi_{-i}^\sigma(I)} \psi(\grave{I}) \\
\vspace{-3mm}
\end{align*}
This gives a bound on the regret at any information set $I$. Taking the regret at the root node gives the desired result.
\end{proof}

This version of our bound weights the error at each information set by the probability of reaching the information set, and similarly, the error at each of the nodes in the information set is weighted by the probability of reaching it. This is important for CFR-style algorithms, where the regret at each information set $I$ only goes to zero when weighted by $\pi^{\sigma}_{-i}(I)$
.
If one wishes to compute an abstraction that minimizes the bound independently of a specific strategy profile, it is possible to take the maximum over all player actions. Importantly, this preserves the probability distribution over errors at nature nodes. In the previous CFR-specific results of \citet{Lanctot12:no-regret}, the reward error bound for each information set was the maximum reward error at any leaf node. Having the reward error be a weighted sum over the nature nodes and only maximized over player action sequences allows significantly finer-grained measurement of similarity between information sets. Consider any poker game where an information set represents the hand that the player holds, and consider three hands: a pair of aces $I_A$, pair of kings $I_K$, or pair of twos $I_2$. When the reward error is measured as the maximum over nodes in the information set, $I_A$ and $I_K$ are as dissimilar as $I_A,I_2$, since the winner changes for at least one hand held by the opponent for both information sets. In contrast to this, when reward errors are weighted by the probability of them being reached, we get that $I_A$ and $I_K$ are much more similar than $I_A$ and $I_2$.

Our proof techniques have their root in those of \citet{Kroer14:Extensive-Form}. We devise additional machinery, mainly Proposition~\ref{pro:bounded-regret-action} and the notion of \gameclass\ abstractions, to deal with  imperfect recall. In doing so, our bounds get a linear dependence on height for the reward approximation error. The prior bounds \citep{Kroer14:Extensive-Form} have no dependence on height for the reward approximation error, and are thus tighter for perfect-recall abstractions.

In general, it is known that imperfect-recall games are harder to solve than perfect-recall games: In the two-player zero-sum case, the problem is NP-hard~\cite{Koller92:Complexity}. However, our game class, \gameclass\ games, is not so broad that it encompasses the type of game used in the proof by \citet{Koller92:Complexity}. More importantly, we are not necessarily interested in solving the imperfect-recall game. Ultimately, we wish to find a strategy that we can map back to the original game, and get a good strategy. 
Theorem~\ref{the:refinement-regret-bound-1} shows that we do not need to solve the imperfect-recall game; we can just compute a strategy with low regret at each information set. To do this, we can employ the CFR algorithm, similar to \citet{Lanctot12:no-regret}.

We now show a second result, which concerns the mapping of Nash equilibria in \gameclass\ games to approximate Nash equilibria in perfect-recall refinements.
\begin{theorem}
For any \gameclass\ game $\Gamma^{\prime}$ and Nash equilibrium $\sigma$,
$\sigma$ is an $\epsilon$-Nash equilibrium when implemented
in any perfect-recall refinement $\Gamma$, where $\epsilon=\max_{i\in N}\epsilon_{i}$
and
\begin{align*}
\vspace{-1mm}
  \epsilon_{i}= &
\max_{\vec{a} \in X_{i}^b(r)}   \sum_{j\in\mathcal{H}_{i}} \sum_{I \in \mathcal{D}_r^{\vec{a},j}} \pi_{-i}^\sigma(I)  \left(  \max_{\breve{I}\in\mathcal{P}(f_{I})} 2 \sum_{s\in I} \frac{\pi^{\sigma}(s)}{\pi^{\sigma}(I)}   \left( \epsilon_{I,\breve{I}}^{0}(s) + \epsilon_{I,\breve{I}}^{R}(s) \right) + \epsilon_{I,\breve{I}}^{D} \right).
\end{align*}
\label{the:refinement-regret-bound-2}
\vspace{-2mm}
\end{theorem}
\begin{proof}
  Assume that we are given a strategy $\sigma$ that is a Nash equilibrium in $\Gamma^\prime$, and a strategy $\sigma^*=(\sigma_{-i},\sigma_i^*)$ where Player $i$ best responds in $\Gamma$. For information sets $I^\prime$ where $\sigma_{-i}(I^\prime)>0,\sigma(I^\prime)=0$, a Nash equilibrium does not put any constraints on behavior. However, we know that Player $i$ could have played a strategy  that satisfies the self-trembling property. Assume any such strategy $\sigma^{ST}$, where it is equal to $\sigma$ everywhere except at such information sets, where a utility-maximizing strategy is played for some arbitrary, fixed distribution over $\mathcal{P}(I^\prime)$. We can then apply Theorem~\ref{the:refinement-regret-bound-1} to get the following (where $\psi(I)$ is defined as in Equation~\ref{equ:psi}):
  \begin{align*}
    \vspace{-2mm}
    V^{\sigma^{*}}_i(r) \leq V^{\sigma^{ST}}_i(r)  +\max_{\vec{a} \in X_{i}^b(r)}   \sum_{j\in\mathcal{H}_{i}} \sum_{\grave{I} \in \mathcal{D}_{r}^{\vec{a},j}} \frac{\pi_{-i}^\sigma(\grave{I})}{\pi_{-i}^\sigma(I)} \psi(\grave{I})
    \vspace{-2mm}
  \end{align*}
Where all regrets $r(I^\prime,a^*)=0$ since $\sigma^{ST}$ is a Nash equilibrium. Now, we observe that the utility is the same for $\sigma$ and any $\sigma^{ST}$ at the root node, $V^{\sigma^{ST}}_i(r) = V^{\sigma}_i(r)$:
  \begin{align*}
    V^{\sigma^{*}}_i(r) \leq V^{\sigma}_i(r)  +\max_{\vec{a} \in X_{i}^b(r)}   \sum_{j\in\mathcal{H}_{i}} \sum_{\grave{I} \in \mathcal{D}_{r}^{\vec{a},j}} \frac{\pi_{-i}^\sigma(\grave{I})}{\pi_{-i}^\sigma(I)} \psi(\grave{I})
    \vspace{-2mm}
  \end{align*}
  which is the result we wanted.
\end{proof}
For practical game solving, Theorem~\ref{the:refinement-regret-bound-1} has an advantage over Theorem~\ref{the:refinement-regret-bound-2}: any algorithm that provides guarantees on immediate counterfactual regret in imperfect-recall games can be applied. For example, the CFR algorithm can be run on a {\gameclass} abstraction, and achieve the bound in Theorem~\ref{the:refinement-regret-bound-1}, with the information set regrets $\pi_{-i}^\sigma(I) r(f_I)$ decreasing at a rate of $O(\sqrt{(T)})$. Conversely, no good algorithms are known for computing Nash equilibria in imperfect-recall games.

\section{Complexity and algorithms} \label{sec:computational-results}
We now investigate the problem of computing \gameclass\ abstractions with minimal error bounds. First, we show that this is hard, even for games with a single player and a game tree of height two.\footnote{\citet{Sandholm12:Lossy} already showed hardness of computing an optimal abstraction when minimizing the actual loss of a unique equilibrium.}
\begin{theorem}
  Given a perfect-recall game and a limit on the number of information sets, determining whether a \gameclass\  abstraction with a given bound as in Theorem~\ref{the:refinement-regret-bound-1} or~\ref{the:refinement-regret-bound-2} exists is NP-complete. 
  This holds even if there is only a single player, and the game tree has height two. 
  \label{the:np-hard}
\end{theorem}

The hardness proof (given in Appendix~\ref{app:hardness}) is by reduction from clustering, which also hints that clustering techniques could be used in an abstraction algorithm within our framework.
Performing abstraction at a single level of the game tree 
that minimizes our bound reduces to clustering if the information sets considered for clustering satisfy Conditions~\ref{con:opponent-information-structure} and~\ref{con:player-information-structure}. The distance function for clustering depends on how the trees match on utility and nature error, and the objective function depends on the topology higher up the tree.
In such a setting, an imperfect-recall abstraction with solution quality bounds can be computed by clustering valid information sets level-by-level in a bottom-up fashion. In general, a level-by-level approach has no optimality guarantees, as some games allow {\em no} abstraction unless coupled with other abstraction at different levels (a perfect-recall abstraction example of this is shown by \citet{Kroer14:Extensive-Form}). However, considering all levels simultaneously is often impossible in practice.
A medical example of a setting where a level-by-level scheme would work well is given by \cite{Chen12:Tractable}, where an opponent initially chooses a robustness measure, which impacts nature outcomes and utility, but not the topology of the different subtrees. Similarly, the {\em die-roll poker} game introduced by \citet{Lanctot12:no-regret} as a game abstraction benchmark is amenable to this approach.

We now show that single-level abstraction problems (SLAPs) where Conditions~\ref{con:opponent-information-structure} and~\ref{con:player-information-structure} of Definition~\ref{def:gameclass} are satisfied for all merges form a metric space together with the distance function that measures the error bound for merging information set pairs.
Clustering problems over metric spaces are often computationally easier, yielding constant-factor approximation algorithms~\cite{Gonzalez85:Clustering,Feder88:Optimal}.
\begin{definition}
  A \emph{metric space} is a set $M$ and a distance function $d:M\times M \rightarrow \mathbb{R}$ such that the following holds for all $x,y,z\in M$:
  \begin{inparaenum}[(\bgroup\bfseries a\egroup)]  
  \item $d(x,y) \geq 0$ 
  \label{con:distance-metric-non-negative}
  \item $d(x,y) =0 \Leftrightarrow x=y$ (identity of indiscernibles) \label{con:distance-metric-identity-of-indiscernibles}
  \item $d(x,y)=d(y,x)$ (symmetry) \label{con:distance-metric-symmetry}
  \item $d(x,y) \leq d(x,z) + d(z,y)$ (triangle inequality) \label{con:distance-metric-triangle-inequality}
  \end{inparaenum}.
  \label{def:metric-space}
\vspace{-1mm}
\end{definition}
\begin{proposition}
  For a set of information sets $\mathcal{I}^m$ such that any partitioning of  $\mathcal{I}^m$ yields a \gameclass\ abstraction (with no scaling, i.e. $\delta_{I,\breve{I}}=1,\forall I,\breve{I}\in \mathcal{I}^m$), and a function $d:\mathcal{I}^m\times \mathcal{I}^m\rightarrow \mathbb{R}$ describing the loss incurred in the error bound when merging $I,\breve{I} \in \mathcal{I}^m$, the pair $\left( \mathcal{I}^m, d \right)$ forms a metric space.
  \label{pro:abstraction-distance-metric}
\end{proposition}
The proof is shown in Appendix~\ref{app:distance_metric}.
Conversely to our result above, if the scaling variables can take on any value, the triangle inequality does not hold, so $\left(\mathcal{I}^m,d\right)$ is not a metric space. 

Consider three information sets $I_1,I_2,I_3$, each with two nodes reached with probability $0.9$ and $0.1$, respectively. Let there be one action at each information set, leading directly to a leaf node in all cases. Let $I_1=\{1,2\},I_2=\{5,11\},I_3=\{10,23\}$, where the name of the node is also the payoff of Player 1 at the node's leaf.  We have that $I_1$ and $I_2$ map onto each other with scaling variable $\delta_{I_1,I_2}=5$ to get $\epsilon_{I_1,I_2}^R=1$ and $I_2,_3$ with $\delta_{I_2,I_3}=2,\epsilon^R_{I_2,I_3}=1$. However, $I_1$ and $I_3$ map onto each other with $\delta_{I_1,I_3}=10$ to get $\epsilon^R_{I_1,I_3}=3$ which is worse than the sum of the costs of the other two mappings, since all reward errors on the right branches are multiplied by the same probability $0.1$, i.e., $0.1 \cdot \epsilon_{I_1,I_2}^R + 0.1 \cdot \epsilon^R_{I_2,I_3} < 0.1 \cdot \epsilon^R_{I_1,I_3}$.

The objective function for our abstraction problem has two extreme versions. The first is when the information set that is reached depends entirely on players not including nature. In this case, the error bound over the abstraction at each level is the maximum error of any single information set. This is equivalent to the minimum diameter clustering problem, where the goal is to minimize the maximum distance between any pair of nodes that share a cluster; \citet{Gonzalez85:Clustering} gave a $2$-approximation algorithm when the distance function satisfies the triangle inequality.
Coupled with Proposition~\ref{pro:abstraction-distance-metric} this gives a 2-approximation algorithm for minimizing our bound on SLAPs.
The other extreme is when each of the information sets being reached differ only in nature's actions. In this setting, the error bound over the abstraction is a weighted sum of the error at each information set. This is equivalent to clustering where the objective function being minimized is the weighted sum over all elements, with the cost of each element being the maximum distance to any other element within its cluster. To our knowledge, clustering with this objective function has not been studied in the literature, even when the weights are uniform. 
Generally, the objective function can be thought of as a tree, where a given leaf node represents some information set, and takes on a value equal to the maximum distance to any information set with which it is clustered. Each internal node either takes the maximum or weighted sum of its child-node errors. The goal is to minimize the error at the root node.
In practice, integer programs (IPs) have sometimes been applied to clustering information sets for EFG abstraction~\cite{Gilpin07:Better,Gilpin07:Potential} (without bounds on solution quality, and just for perfect-recall abstractions), and are likely to perform well in our setting. An IP can easily be devised for any objective function in the above form. 

For abstraction problems across more than a single level, Proposition~\ref{pro:abstraction-distance-metric} does not give any guarantees. While the result can be applied level-by-level, the abstraction performed at one level affects which information sets are valid for merging at other levels, and thus the approximation factor is not preserved across the levels.


\section{Experiments}
We now investigate what the optimal SLAP bounds (in terms of Theorem~\ref{the:refinement-regret-bound-2}) look like for the \emph{die roll poker (DRP)} game, a benchmark game for testing abstraction \citep{Lanctot12:no-regret}.
Die-roll poker is a simple two-player zero-sum poker game where dice, rather than cards, are used to determine winners. At the beginning of the game, each player antes one chip to the pot.
The game then consists of two rounds. In each round, each player rolls a private die (making the game imperfect information). Afterwards a betting round occurs.
During betting rounds, a player may fold (causing the other player to win), call (match the current bet), or raise by a fixed amount, with a maximum of two raises per round. In the first round, each raise is worth two chips. In the second round, each raise is worth four chips. The maximum that a player can bet is $13$ chips, if each player uses all their raises. At the end of the second round, if neither player has folded, a showdown occurs. In the showdown, the player with the largest sum of the two dice wins all the chips in the pot. If the players are tied, the pot is split.

\begin{figure}[]
  \begin{center}
    \includegraphics[scale=0.40]{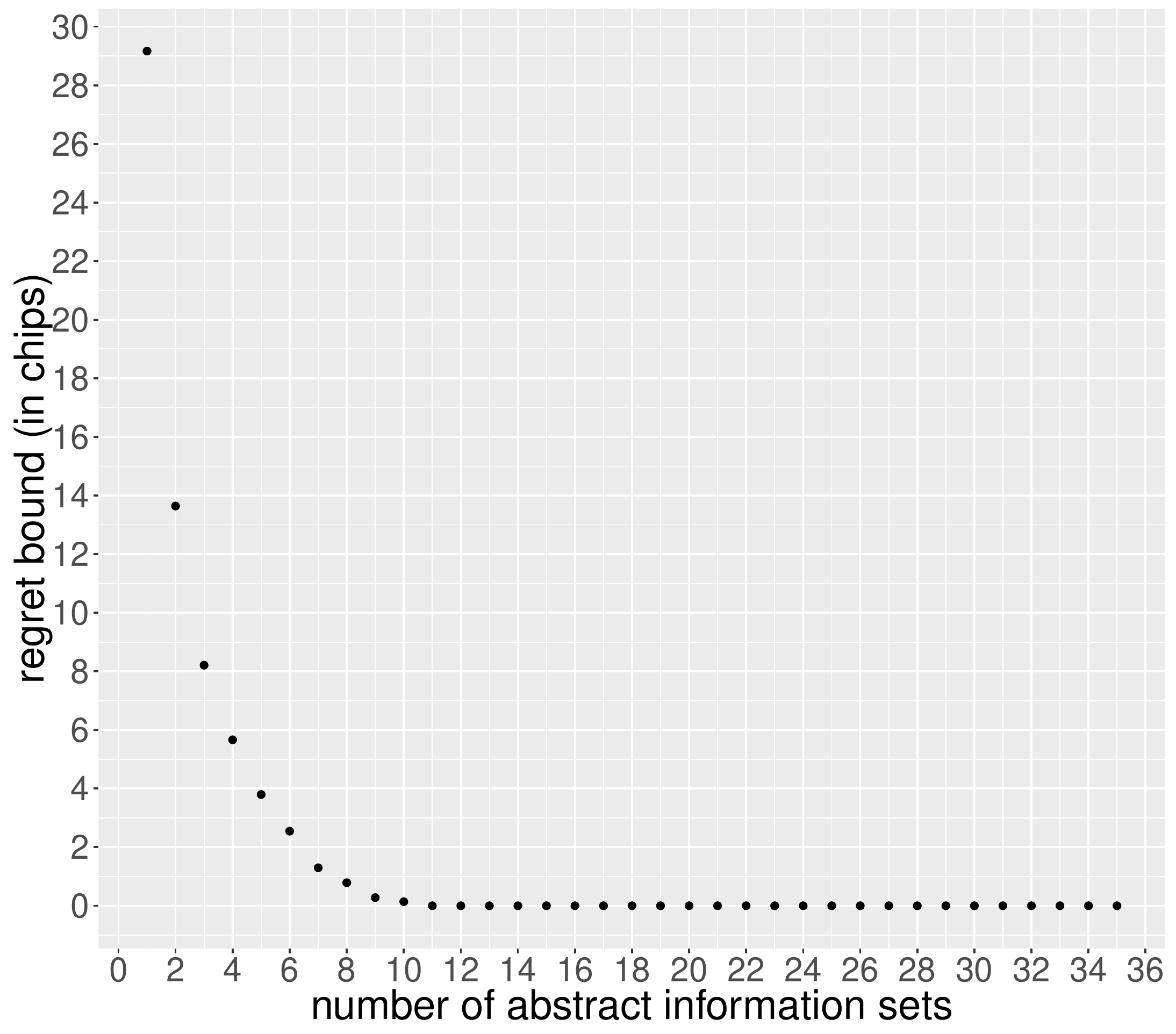}
  \end{center}
  \figreduc
  \caption{Regret bounds for varying SLAP sizes in DRP. The x-axis shows the number of information sets in the abstraction, and the y-axis shows the theoretical bound on soluation quality. The total number of information sets in the original game is $36$}
  \label{fig:experimental_results}
  \figreduc
\end{figure}

DRP has the nice property that abstractions computed at the bottom level of the tree satisfy the conditions of Definition~\ref{def:gameclass}. At heights above that one we can similarly use our clustering approach, but where two information sets are eligible for merging only if there is a bijection between their future die rolls such that the information sets for the future rolls in the bijection have been merged. A clustering would be computed for each set in the partition that represents a group of information sets eligible for merging. In the experiments in this paper we will focus on abstraction at the bottom level of the tree. We use CPLEX to solve an IP encoding the SLAP of minimizing our bound given a limit on the number of abstract information sets. 
The results are shown in Figure~\ref{fig:experimental_results}.
For one or two clusters, the bound is bigger than the largest payoff in the game, but already at three clusters it is significantly lower. At eight clusters, the bound is smaller than that of always folding, and decreases steadily to zero at eleven clusters (the original game has 36 information sets). While these experiments show that our bound is relatively small for the DRP game, they are limited in that we only performed abstraction at a single level. If abstraction at multiple levels is performed, the bound is additive in the error over the levels.

Another important question is how well strategies computed in abstractions that are good---as measured by our bound---perform in practice. \citet{Lanctot12:no-regret} conducted experiments to investigate the performance of CFR strategies computed in imperfect-recall abstractions of several games: DRP, Phantom tic-tac-toe (where moves are unobserved), and Bluff. They found that CFR computes strong strategies in imperfect-recall abstractions of all these games, even when the abstraction did not necessarily fall under their framework. Their experiments validate a subset of the class of \gameclass\ abstractions: ones where there is no nature error. 
Due to this existing experimental work, we focus our experiments on problems where abstraction does introduce nature error. One class of problems where such error can occur are settings where players observe imperfect signals of some phenomenon. For such settings, one would expect that there is correlation between the observations made by the players. Examples include negotiation, sequential auctions, and strategic acquisition.

DRP can be thought of as a game where the die rolls are the signals. Regular DRP has a uniform distribution over the signals. We now consider a generalization of DRP where die rolls are correlated: {\em correlated die-roll poker} (CDRP). There are many variations on how one could make the rolls correlated; we use the following. We have a single correlation parameter $c$, and the probability of any pair of values $(v_1,v_2)$, for Player $1$ and $2$ respectively, is $\frac{1}{\#sides^2} - c\left| v_1 - v_2 \right|$. The probabilities for the second round of rolls is independent of the first round. As an example, the probability of Player~$1$ rolling a $3$ and Player$2$ rolling a $5$ with a regular $6$-sided die in either round would be $\frac{1}{36}-2c$. We generate DRP games with a $4$-sided die and $c \in \{ 0, 0.01, 0.02, 0.03, 0.04, 0.05, 0.06, 0.07 \}$.

\begin{figure}[]
  \begin{center}
    \includegraphics[scale=0.475]{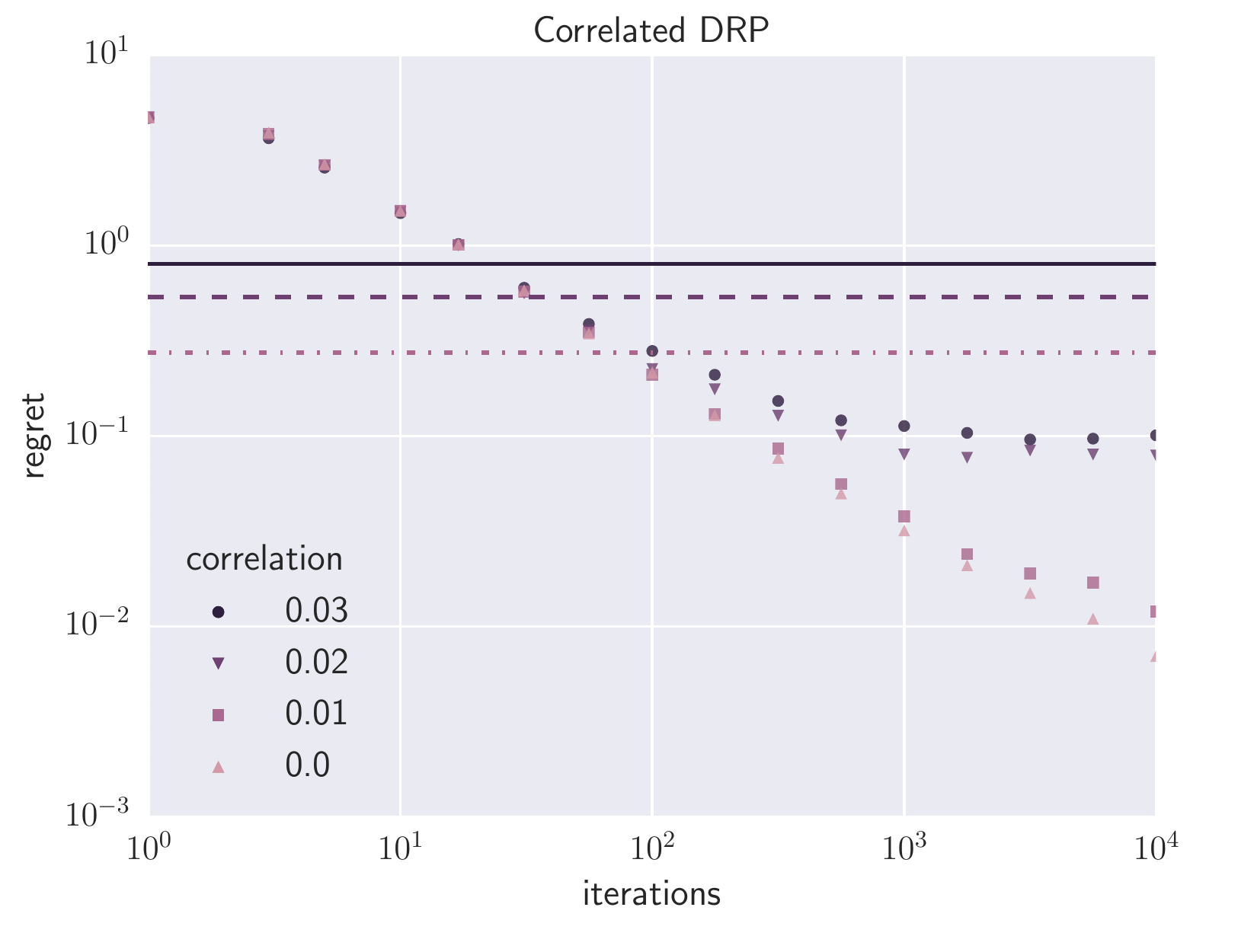}
    \includegraphics[scale=0.475]{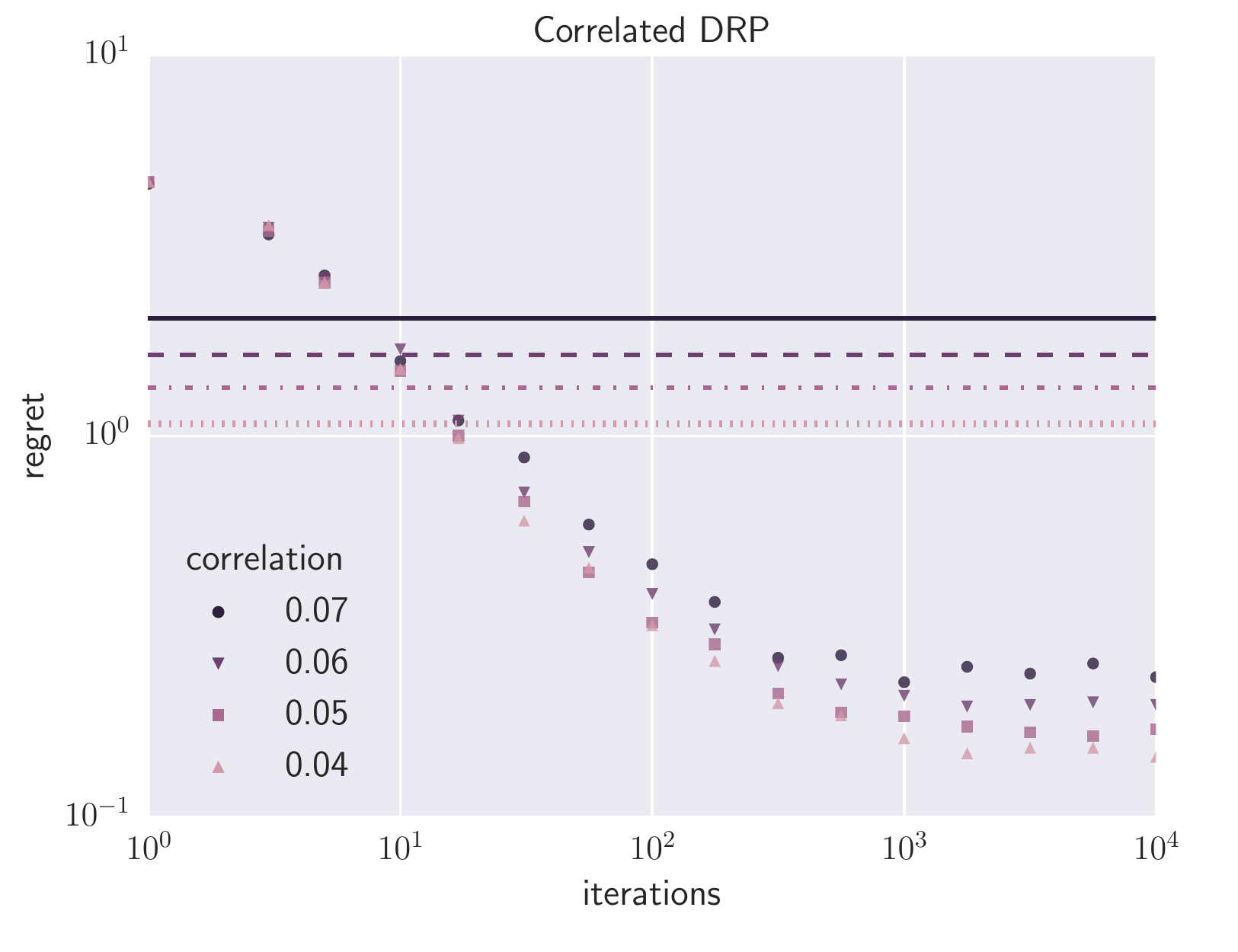}
  \end{center}
  \figreduc
  \caption{Log-log plots of the sum of the two players' regrets as a function of CFR iterations on the bound-minimizing abstraction of CDRP. The legends give the amount of correlation in the die rolls of the different CDRP games on which we ran experiments. The horizontal lines show the respective ex-ante regret bound of Theorem~\ref{the:refinement-regret-bound-2} for each of the CDRP games. (In the first game on the left where the correlation is zero, the abstraction is lossless, so the horizontal line (not shown) would be at zero.)}
  \label{fig:cfr_results}
  \figreduc
\end{figure}

For each value of $c$, we compute the optimal bound-minimizing abstraction for the second round of rolls, with a static mapping between information sets such that for any sequence of opponent rolls,
the nodes representing that sequence in either information set are mapped to each other. The bound cost of the mappings is precomputed, and the optimal abstraction is found with a standard MIP formulation of clustering. 
After computing the optimal abstraction for a given game, we run CFR on the abstraction, and measure the regret for either player in terms of their regret in the full game. Figure~\ref{fig:cfr_results} shows the results of these experiments. On the x-axis is the number of CFR iterations. On the y-axis is $r_1+r_2$, where $r_i$ is the regret for Player~$i$ for the strategy at a given iteration. Furthermore, the horizontal lines denote the regret bound of Theorem~\ref{the:refinement-regret-bound-2} for an exact Nash equilibrium.
On the left in Figure~\ref{fig:cfr_results} is shown the results for the four smallest values of $c$, on the right the four largest values. As can be seen, CFR performs well on the abstractions, even for large values of $c$: when $c=0.7$, a very aggressive abstraction, the sum of regrets still goes down to $\sim0.25$ (for reference, always folding has a regret of $1$). We also see that for $c\geq 0.2$, the regret stops decreasing after around $1000$ iterations. This is likely where CFR converges in the abstraction, with the remaining regret representing the information lost through the abstraction. We also see that our theoretical bound is at the same order of magnitude as the actual bound even when CFR converges.

\section{Discussion}
In this paper, we proved bounds for abstractions obtained through merging information sets.
The perfect-recall results of \citet{Kroer14:Extensive-Form} also allow abstraction by removing actions available to players.
The following approach can be adopted for imperfect-recall abstraction with such branch removal, while still obtaining solution-quality guarantees. First, a valid perfect-recall abstraction is computed, where the desired branches are removed. The results by \citet{Kroer14:Extensive-Form} give bounds on the solution quality of equilibria computed in this abstraction. An imperfect-recall abstraction can then be computed from this perfect-recall abstraction, with our results providing bounds on solution quality for this step. Solution quality bounds can then be achieved for the final abstraction by taking the sum of the bounds for the two steps.
It is likely that tighter bounds could be derived by analyzing the distance between the original game and the final abstraction directly. 
We leave this as future research.

An important avenue for future research is how to merge the solution-quality bounds obtained in this work with practical algorithms for generating abstractions. We showed how single-level abstraction problems can be addressed. For multi-level abstraction, a similar approach can be adopted, but where the abstraction is either computed greedily level-by-level, or using IP or search algorithms that ensure that the abstraction satisfies Conditions 1 and 2 of Definition~\ref{def:gameclass} across levels. \citet{Kroer15:Discretization} showed that the results for perfect-recall abstraction from \citet{Kroer14:Extensive-Form} can be used to prove bounds for {\efg}s with continuous action spaces. It would be interesting to show similar results for the imperfect-recall setting.



\begin{small}
  \bibliographystyle{acmsmall}
  \bibliography{dairefs}
\end{small}

\pagebreak
\appendix

\section{Proof of Theorem~\ref{the:np-hard}}
\label{app:hardness}
\begin{proof}
  Consider the {\em two-dimensional k-center clustering decision problem} with the $L_q$ distance metric.
  It is defined as follows: given a set $P=\{(x_1,y_1),\ldots,(x_n,y_n)\}$ of $n$ points in the plane, and an integer $k$, does there exist a partition of $P$ into $k$ clusters $\mathcal{C}=\{c_1,\ldots, c_k\}$ such that the maximum distance $\| p - p' \|_q \leq c$ between any pair of points $p,p'$ in the same cluster is minimized.
  This problem is NP-hard to approximate within a factor of $2$ for $q=\infty$, amongst others. \cite{Feder88:Optimal}.

  Given such a problem, we construct a perfect-recall game as follows. For each point $p\in P$, we construct an information set $I_p$. We insert two nodes $s_p^x,s_p^y$ in each information set $I_p$, representing the dimensions $x,y$ respectively. All these nodes descend directly from the root node $r$, where
  Player $1$ acts. At each information set we have two actions, $a_c,a_v$. For any point $p$, we add leaf nodes at the branch $a_c$ with payoff $M,2M$ at the nodes $s_p^x,s_p^y$ respectively. If we pick a sufficiently large $M$, this ensures that for any two points $p,p'$, their nodes $s_p^x,s_{p'}^x$ will map to each other, and similarly for $y$. This also ensures that the scaling variable has to be set to $1$ for all information set mappings.
  For the branches $a_v$, we add leaf nodes with utility equal to the $x,y$  coordinate of $p$ at the $s_p^x,s_p^y$ nodes respectively.

  There is a one-to-one mapping between clusterings of the points $P$ and partitions of the information sets $\{I_p:p\in P\}$. The quality of a clustering is
  $\max_{z \in \{x,y\}} \max_{j=1,\ldots,k} \max_{p,p' \in c_j} \left| p(z) - p'(z) \right|$.
   Since Player $1$ acts at $r$, the abstraction quality bound is equal to the maximum difference over any two leaf nodes mapped to each other, as $\epsilon^0=\epsilon^D = 0$. This is the same as the quality measure of the clustering. Thus, an optimal $k$ size clustering is equivalent to an optimal $k$ information set abstraction.

 Given some \gameclass\ abstraction, verifying the solution is easy to do: in one top-down traversal of the game tree, compute the node distributions at each information set. For each full-game information set, this gives the distribution-approximation error. For each information set pair mapped to each other, the transition- and reward-approximation error can now be computed by a single traversal of the two. Thus the problem is in NP.
\end{proof}

\section{Proof of Proposition~\ref{pro:abstraction-distance-metric}}
\label{app:distance_metric}
\begin{proof}
  The first condition follows from the other three. Condition~\ref{con:distance-metric-identity-of-indiscernibles}, identity of indiscernibles, does not hold for information sets. However, any pair of information sets with distance zero can be merged losslessly in preprocessing, thus rendering the condition true (having distance zero is transitive, so the minimal preprocessing solution is unique). Condition~\ref{con:distance-metric-symmetry}, symmetry, holds by definition, since our distance metric is defined as the error incurred from merging two information sets, which considers the error from both directions of the mapping.

  Finally, we show that Condition~\ref{con:distance-metric-triangle-inequality}, the triangle inequality holds.
Consider any three information sets $I_{1},I_{2},I_{3}\in\mathcal{I}^{m}$.
We need to show that $d(I_{1},I_{3})\leq d(I_{1},I_{2})+d(I_{2},I_{3})$.
Let $\phi_{I_{1},I_{2}},\phi_{I_{2},I_{3}}$ be the mappings for $I_{1},I_{2}$
and $I_{2},I_{3}$ respectively. We construct a mapping $\phi_{I_{1},I_{3}}=\phi_{I_{2},I_{3}}\circ\phi_{I_{1},I_{2}}$
and show that it satisfies the triangle inequality. For the leaf payoff
error, since $\delta_{I_{1},I_{2}}=\delta_{I_{2},I_{3}}=1$, at any
leaf $z\in Z_{I_{1}}$ we get:
\[
u_{i}(z)\leq u_{i}(\phi_{I_{1},I_{2}}(z))+\epsilon_{I_{1},I_{2}}(z)\leq u_{i}(\phi_{I_{2},I_{3}}(\phi_{I_{1},I_{2}}(z))) +\epsilon_{I_{2},I_{3}}(\phi_{I_{1},I_{2}}(z))+\epsilon_{I_{1},I_{2}}(z)
\]
For the nature leaf probability error we can apply the same reasoning:
\begin{align*}
 & \pi^\sigma_{0}(z[I_{1}], z)\\
\leq{} & \pi^\sigma_{0}(\phi_{I_{1},I_{2}}(z[I_{1}]), \phi_{I_{1},I_{2}}(z))+\epsilon_{I_{1},I_{2}}^{0}\\
\leq{} & \pi^\sigma_{0}(\phi_{I_{2},I_{3}}(\phi_{I_{1},I_{2}}(z[I_{1}])), \phi_{I_{2},I_{3}}(\phi_{I_{1},I_{2}}(z)))+\epsilon_{I_{2},I_{3}}^{0}(\phi_{I_{1},I_{2}}(z))+\epsilon_{I_{1},I_{2}}^{0}(z)
\end{align*}
Again, we derive the distribution error using a similar approach:
\begin{align*}
 & \frac{\pi_{0}(z[I_{1}])}{\pi_{0}(I_{1})}\\
\leq{} & \frac{\pi_{0}(\phi_{I_{1},I_{2}}(z[I_{1}]))}{\pi_{0}(I_{2})}+\epsilon_{I_{1},I_{2}}^{D}(z[I_{1}])\\
\leq{} & \frac{\pi_{0}(\phi_{I_{2},I_{3}}(\phi_{I_{1},I_{2}}(z[I_{1}])))}{\pi_{0}(I_{3})}+\epsilon_{I_{2},I_{3}}^{D}(\phi_{I_{2},I_{3}}(z[I_{1}]))+\epsilon_{I_{1},I_{2}}^{D}(z[I_{1}])
\end{align*}
This completes the proof.
\end{proof}



\end{document}